%% file: main-sigconf.tex
\documentclass[sigconf]{acmart}

\AtBeginDocument{%
  \providecommand\BibTeX{{%
    \normalfont B\kern-0.5em{\scshape i\kern-0.25em b}\kern-0.8em\TeX}}}

\copyrightyear{2020}
\acmYear{2020}
\setcopyright{acmcopyright}

\acmConference[CIKM '20]{The 29th ACM International Conference on Information and Knowledge Management}{October 19--23, 2020}{Virtual Event, Ireland}
\acmBooktitle{The 29th ACM International Conference on Information and Knowledge Management (CIKM '20), October 19--23, 2020, Virtual Event, Ireland}
\acmPrice{15.00}
\acmDOI{10.1145/3340531.3411901}
\acmISBN{978-1-4503-6859-9/20/10}

\acmSubmissionID{fp0451}

\usepackage{booktabs} 

\usepackage{xspace}
\usepackage{xcolor}
\usepackage{amsfonts}
\usepackage{subfigure}
\usepackage{multirow}
\usepackage[inline]{enumitem}

\newcommand{\palpha}{\ensuremath{\text{P}^3\!\alpha}\xspace}

\newcommand{\pbeta}{\ensuremath{\text{RP}^3\!\beta}\xspace}

\newcommand{\iALS}{iALS\xspace}

\newcommand{\idest}{i.e.,\xspace}

\newcommand{\ConvNCF}{ConvNCF\xspace}
\newcommand{\CFM}{CFM\xspace}
\newcommand{\CoupledCF}{CoupledCF\xspace}
\newcommand{\BPRMF}{MF BPR\xspace}
\newcommand{\MF}{matrix-factorization\xspace}

\begin{document}
\fancyhead{} 
\title{Critically Examining the Claimed Value of Convolutions over User-Item Embedding Maps for Recommender Systems}

\author{Maurizio Ferrari Dacrema}
\orcid{0000-0001-7103-2788}
\affiliation{%
  \institution{Politecnico di Milano, Italy}
}
\email{maurizio.ferrari@polimi.it}

\author{Federico Parroni}
\affiliation{%
  \institution{Politecnico di Milano, Italy}
}
\email{federico.parroni@mail.polimi.it}

\author{Paolo Cremonesi}
\orcid{0000-0002-1253-8081}
\affiliation{%
  \institution{Politecnico di Milano, Italy}
}
\email{paolo.cremonesi@polimi.it}

\author{Dietmar Jannach}
\orcid{0000-0002-4698-8507}
\affiliation{%
  \institution{University of Klagenfurt, Austria}
}
\email{dietmar.jannach@aau.at}


\begin{abstract}
In recent years, algorithm research in the area of recommender systems has shifted from matrix factorization techniques and their latent factor models to neural approaches. However, given the proven power of latent factor models, some newer neural approaches incorporate them within more complex network architectures.
One specific idea, recently put forward by several researchers, is to consider potential correlations between the latent factors, i.e., embeddings, by applying convolutions over the user-item interaction map. However, contrary to what is claimed in these articles, such interaction maps do not share the properties of images where Convolutional Neural Networks (CNNs) are particularly useful.
In this work, we show through analytical considerations and empirical evaluations that the claimed gains reported in the literature cannot be attributed to the ability of CNNs to model embedding correlations, as argued in the original papers. Moreover, additional performance evaluations show that all of the examined recent CNN-based models are outperformed by existing non-neural machine learning techniques or traditional nearest-neighbor approaches. On a more general level, our work points to major methodological issues in recommender systems research.
\end{abstract}

\ccsdesc[500]{Information systems~Recommender systems}
\ccsdesc[500]{Computing methodologies~Neural networks}

\keywords{Deep Learning; Evaluation; Convolutional Neural Networks}

\maketitle

\input{main-body.tex}

\bibliographystyle{ACM-Reference-Format}
\bibliography{main-references}

\end{document}

%% file: main-body.tex
\section{Introduction}
In the era of exponential information growth, recommender systems have proven to be valuable tools to help users explore the vast number of options at their disposal.
Over the last two decades, a large number of collaborative filtering algorithms was proposed for item ranking and relevance prediction, from early nearest-neighbors approaches to machine learning models \cite{Resnick:1994:GOA:192844.192905,sarwar2001item,hu2008IALS,rendle2009bprMF,paudel2017Rp3beta}.
Among these techniques, \emph{\MF} (MF) algorithms were particularly popular in the past decade after the Netflix Prize competition \cite{bennett2007netflix}, both in industry and academia. These algorithms project users and items into a low dimensional latent space \cite{Koren2011,johnson2014logistic}, and an interaction between a user and an item is usually modeled as the dot product of their respective latent vectors.

In recent years, the attention of researchers and industry has moved to deep learning (neural) approaches for collaborative filtering, and various network architectures have been proposed, e.g., \cite{Zhang:2019:DLB:3309872.3285029,CovingtonYouTube2016,he2017neural}.
In several of these approaches, researchers try to incorporate specific architectural components into their models that were previously found to be effective in other application domains of deep learning. Examples of such architectural components are attention layers, autoencoders and convolution layers \cite{Zhang:2019:DLB:3309872.3285029,liang2018variationalautoencodersforCF}.
In particular Convolutional Neural Networks (CNNs) were successfully applied to different recommendation-related tasks, including image or text feature extraction for content-based models \cite{van2013deep}, sequential recommendations \cite{tang2018personalized} or collaborative filtering \cite{Zhang:2019:DLB:3309872.3285029}.

In some recent works, different proposals were also put forward to combine the proven power of low dimensional approximation models (e.g., latent factor models) with CNNs.
In particular, one underlying idea of the proposals in \cite{ConvNCFIJCAI2018,Du:2019:MED:3357218.3357154,ijcai2019CFM,Zhang2018CoupledCF} is to use CNNs to model and leverage correlations in the latent factors (embeddings) space.
In all these papers, the respective authors claim to have obtained significant gains in accuracy by applying convolutions over user-item interaction maps derived from the outer product of user-item embeddings.

In some of these works---all of them originally published at the highly-ranked IJCAI conference---the authors argue that these interaction maps can be viewed as analogous to images, an application area in which CNNs are very effective.
However, user-item interaction maps, as produced by common
latent space approaches, do not share the properties of images. For most common latent space models, there is no natural order of the elements in the latent vectors, the dimensions are independent, and the correlation between the latent dimensions is actually not modeled. Therefore, it is more than surprising that the above-mentioned CNNs approaches were able to benefit from detecting correlations between the dimensions.

In this work, we therefore critically examine the progress claimed in these papers, based on both theoretical considerations and experimental evaluations. In particular, we report the results of ablation studies that were not present in the original papers. These results indicate that the proposed CNN-based models do \emph{not} capture correlations between the embedding dimensions as claimed in the original papers. Rather, they merely act as a non-linear function of their element-wise product and the removal of the embedding correlations leads to no significant effects.
Our work therefore points to major issues in the way these articles have justified and demonstrated their claims. A particular problem in that context may lie in the \emph{missing validation of the assumptions} these models rely upon.\footnote{A clear explanation of such underlying assumptions is also stipulated in the \emph{Machine Learning Reproducibility Checklist} (\url{https://www.cs.mcgill.ca/~jpineau/ReproducibilityChecklist-v2.0.pdf}), which was, for example, adopted in recent years as part of the NeurIPS submission process.}

The question however still remains how the authors were able to demonstrate significant performance gains in their experimental evaluation. One reason might lie in the choice of the baselines that were used in their evaluation. If the baselines were not strong, demonstrating a ``win'' over previous methods might be possible even if the added CNN layer merely acts as an additional universal approximator function. We have therefore conducted additional experiments, using the exact same experimental setup as in the original papers, where we benchmarked the new methods against established non-neural machine learning models and traditional nearest-neighbor techniques. Our results show that for all algorithms, datasets, and metrics, existing techniques outperformed the recent CNN-based methods. We share the code and data used in our experiments online.\footnote{\url{https://github.com/MaurizioFD/RecSys2019_DeepLearning_Evaluation}}

On a more general level, our work adds to the growing evidence that we, as a community, face significant methodological problems, which makes it difficult to judge if we are truly moving the field much forward.
Previous works in the area of recommender systems \cite{DBLP:journals/corr/abs-1905-01395,Ludewig2018,Ferraridacrema2019}, information retrieval \cite{Yang:2019:CEH:3331184.3331340,Armstrong:2009:IDA:1645953.1646031} or time-series forecasting \cite{Makridakis2018} found that the progress achieved with certain complex models is sometimes non-existent and was only observed because the chosen baselines were weak or not properly optimized. 
In another recent paper Rendle et al. \cite{ncfvsmf2020} confirmed previous results that NCF \cite{he2017neural} is not able to consistently outperform non-neural baselines \cite{Ferraridacrema2019,Ferraridacrema2019troubling}, and  showed it is not trivial to learn a dot product with a multilayer perceptron. 
These works indicate that sometimes the experimental evaluations are not well suited to demonstrate the authors' claims regarding which part of a complex architecture actually contributes to an observed performance gain. 
The same observation was recently made by Lipton et al. \cite{troubling-trends-1807.03341}, where it was argued that sometimes papers present complex algorithms involving several components (e.g., architecture, preprocessing steps, training procedure) without reporting any ablation study to clarify the contribution of the individual components. 
The article encouraged authors to ask ``What worked?'' and ``Why?'' rather than just ``How well?'', highlighting the importance of sound empirical inquiry, which can yield new knowledge or insights even when no new algorithm is proposed.

With our study, we address such questions and encounter signs that the lack of theoretical underpinnings of new deep learning approaches may contribute to the observed problems.

\section{Background}
\label{sec:convolution_rs}

\subsection{Principles and Assumptions of CNNs}
A convolutional neural network is a multilayer feed-forward neural network that was originally developed to address image recognition problems \cite{zhang2018integrating}. Unlike other types of neural networks, CNNs were therefore designed for certain types of inputs, e.g., images, which have a specific topology.
As stated in \cite{Krizhevsky2012}, the paper describing the seminal \emph{AlexNet} model for image classification, CNNs are based on two strong and important assumptions regarding the nature of the processed input data (i.e., the images): \emph{stationarity} of statistics and \emph{locality} of pixel dependencies, that is, pixels that are close will be strongly correlated \cite{lecun1998gradient}.

With data exhibiting these properties, local features  (e.g., lines, corners) emerge from their respective immediate surroundings, regardless of their absolute position in the observed data.
CNNs have also been proven effective in several other scenarios where the data exhibits feature locality (e.g., time series forecasting, natural language processing). The importance of translation invariance and feature locality for CNNs has been widely discussed, both in terms of \emph{spatial locality} \cite{Yu2015MultiScaleCA,long2015fully,lee2009convolutional,turaga2010convolutional}, and \emph{time locality} for sequence modeling \cite{BaiTCN2018}.

Technically, CNNs in their traditional form consist of a convolution layer and a pooling layer.\footnote{The use of max-pooling to reduce the dimensionality of the data has been recently criticized, see \cite{sabour2017dynamic}.}
The convolution layer uses a kernel, with certain weights, which is moved across the two dimensional feature matrix (i.e., the image). Sharing the kernel weights across the image allows a CNN to have much fewer parameters than a fully connected NN and to leverage the spatial locality and location invariance of features. The pooling layer reduces the dimensionality of the data, allowing successive convolution layers on a broader field of view. Multiple convolutional layers are then able to interpret increasingly complex patterns by further aggregating lower-level features.
As stated in \cite{lecun1998gradient}, after the detection of a feature, only the relative position of that feature with respect to other features is relevant, not the absolute one.

Convolution is usually applied on a \emph{local area} (i.e., the kernel size) of a two-dimensional map of a certain size. Identifying the \emph{local area} on which to apply the convolution requires a definition of \emph{proximity} between points. Depending on the use case, different definitions of proximity may be used. In case of images, the proximity is defined in spatial terms, i.e., pixels that are close in the image will be perceived as close by an observer and are therefore meaningful for the reconstruction of more complex patterns.
Other definitions of proximity have also been developed for non-Euclidean data like social networks or knowledge graphs \cite{defferrard2016convolutional}.

\subsection{Use of CNNs in Recommender Systems}
A growing number of papers aim to use CNNs for collaborative filtering tasks. Most existing approaches can be grouped into three categories \cite{Zhang:2019:DLB:3309872.3285029}:
\begin{description}
    \item[Feature extraction:] In this case, a CNN is used to extract features from heterogeneous data sources, e.g., images, video, audio, which are then used in another recommendation model \cite{van2013deep}.
    \item[Pretrained embeddings:] In such approaches, a CNN is applied on user or item embeddings that were pretrained by another model, e.g., \cite{ijcai2019CFM,ConvNCFIJCAI2018}.
    \item[Learnable embeddings:] Also in this case the CNN is applied on user or item embeddings. Differently from the previous case, the embeddings are an integral part of the model and are trained along with the CNN (i.e., they are not pre-trained with another approach), e.g., in \cite{Zhang2018CoupledCF,gaussianconv2019}.
\end{description}
In this paper we will focus on the last two cases, when CNNs are applied on embeddings.\footnote{Note that we only consider approaches that use a user-item rating matrix as an input. CNNs were also applied for session-based recommendation \cite{DBLP:conf/wsdm/YuanKAJ019}, where they however showed some limitations as well \cite{LudewigMauro2019}.
}

If we compare images or graph data to embeddings in the specific form of latent factors derived from a user-item rating matrix, there is a key difference.
For images or graphs, there is a ``natural'' way of defining the local area, e.g., based on the distance in pixels or the number of hops in the graph. The corresponding proximity measure has a strong relation with the data semantics.

However, the papers analyzed in this work, i.e., \cite{ConvNCFIJCAI2018,Du:2019:MED:3357218.3357154,ijcai2019CFM,Zhang2018CoupledCF}, do not clearly provide a definition of locality for the embeddings, nor do they  describe the semantic topology of the input data.
In the \ConvNCF approach \cite{ConvNCFIJCAI2018}, for example, the input to the CNN is a user-item interaction map
that is created by computing the outer product of embeddings pretrained using matrix factorization. While the authors argue that this map is analogous to an image and therefore the use of CNN is justified, they do not demonstrate or discuss in detail what the resulting map topology should represent and whether it possesses typical image properties (e.g., spatial locality and translation invariance). 
Technically, the interaction map created by the outer product of the embeddings in the described approach contains two components: (i) the main diagonal that represents the element-wise product between two embedding vectors and (ii) the off-diagonal elements that capture embeddings correlations.
Unfortunately, none of the analyzed papers measured and compared the contribution of the two components to the proposed CNN model. In all papers the authors claim that the CNN is able to model the correlations between embeddings based on the fact that CNN models outperform other models that are not using convolutions. This, we argue, is not sufficient. Comparing models with vastly different structures means that a multitude of factors may have an impact on the results. Ultimately, it is not entirely clear from the provided experiments how much the correlations between embeddings, as modeled by the CNN, actually contributed to the observed performance gains.

\section{Overview of Analyzed Approaches}
In this paper we examine three recent neural approaches that use convolutions over embeddings derived from a user-item rating matrix. All approaches were published at previous \mbox{IJCAI} conferences and for all of them the source code was available.
We identified an additional relevant work \cite{gaussianconv2019}. Its experimental setting (i.e., source code and data) was however not reproducible based on the material provided by the authors.

\subsection{Convolutional Neural Collaborative Filtering}
\label{sec:ConvNCF_model}
Convolutional Neural Collaborative Filtering (\emph{\ConvNCF}) was proposed in \cite{ConvNCFIJCAI2018}. The \ConvNCF model is trained in two steps.
First, a \MF model is fitted on the data.
Then, for each user-item interaction, the outer product of their embedding is computed, resulting in a two-dimensional \emph{interaction map} on which the CNN is applied.

Following the original notation, let $u$ be a user and $i$ an item, $P \in \mathbb{R} ^ {M \times K}$ and $Q \in \mathbb{R} ^ {N \times K}$ the embedding matrix of users and items, respectively; $K$ the embedding size, $M$ the number of users and $N$ the number of items. Lastly, let $p_u,q_i \in \mathbb{R} ^ {K}$ be their respective embeddings. 
Based on these embeddings, the interaction map $E \in \mathbb{R} ^ {K \times K}$ is obtained by computing their outer product as follows:
\begin{align}
E &= p_u \oplus q_i = p_u q_i^T \nonumber\\
\label{eqn:interaction_map}
e_{x,y} &= p_{u,x} \cdot q_{i,y}
\end{align}

In the original paper, pretraining is performed with an \BPRMF model \cite{rendle2009bprMF} and, in a subsequent paper, the authors extend the pretraining to FISM and SVD++ models \cite{Du:2019:MED:3357218.3357154}.
As mentioned before, the interaction map is said to be analogous to an image, but there is no deeper discussion of this claim, which is therefore not verified, and the CNN is said to model embedding correlations but no direct measurement of their contribution is provided.

\subsection{Convolutional Factorization Machines}
\label{sec:CFM_model}
Convolutional Factorization Machines (\emph{\CFM}) were proposed in \cite{ijcai2019CFM} as a context-aware model able to overcome the limited modeling capacity of Factorization Machines (FM) \cite{FactorizationMachines2010}, which are constrained by a linear representation of feature interactions. Similarly to \mbox{\ConvNCF}, \CFM applies a convolution operation on the outer product of the embeddings. In the \CFM approach, however, the embeddings are obtained from a FM via a self-attention layer, which reduces the model's dimensionality. The outer product of the embeddings is computed independently for each context and all interaction maps are stacked on top of each other, forming an interaction cube, on which 3D convolution is applied. If $C$ is the number of contextual features, the interaction maps in the memory cube will be $C(C-1)/2$.

Although the scenario of application is different from \mbox{\ConvNCF} and the embeddings are obtained from a different pretraining step, \CFM has the same theoretical problems as \ConvNCF in that the outer product is treated as if it were an image, without ever demonstrating that this assumption holds. The ability of \CFM to model embeddings interaction is again claimed based on it outperforming baselines with quite different architectures, including \ConvNCF.

\subsection{Coupled Collaborative Filtering}
\label{sec:CoupledCF_model}
Coupled Collaborative Filtering (\emph{\CoupledCF}) was proposed in \cite{Zhang2018CoupledCF} as a method to learn implicit and explicit couplings between users and items, taking advantage of them not being independent, to leverage side information more effectively (e.g., user demographics, item features).
The model is composed of two cooperating architectures, one is a deep collaborative filtering model which only uses user-item interactions, while the other is a CNN on user and item embeddings whose aim is to learn the couplings.

\CoupledCF is different from the other two methods examined here in that the embeddings are not pretrained but are parameters of the model to be learned.
In \cite{Zhang2018CoupledCF}, the authors show experimentally that the quality of the model is improved by adding the CNN and claim this demonstrates that the model is effectively learning the couplings.
Again, as previously mentioned, the paper does not distinguish between the contribution of the the embeddings correlation and the element-wise product.\footnote{An additional issue regarding the choice and optimization of the baselines used in \cite{Zhang2018CoupledCF} was recently reported in \cite{Ferraridacrema2019troubling}.}

\section{Analysis}
\label{sec:critical_analysis}
One fundamental assumption of the analyzed papers is that the interaction map computed via an outer product is analogous to an image (i.e., exhibits spatial locality and translation invariance).
In this section, we demonstrate why this is not the case. We will first discuss this aspect theoretically and then present the results of two empirical studies. In the first study we show that changing the input topology (i.e., the ordering of the latent factors) does not have an impact on accuracy.
Clearly, if the interaction map had local features a significant drop in accuracy would be expected.
The second study consists of two ablation analyses showing that the correlations between embeddings do not provide a statistically significant contribution to the accuracy of the CNN models.

\subsection{Theoretical Considerations}
\label{subsec:theoretical-considerations}
As previously stated, CNNs were developed to model data that exhibits feature locality and a strong topology. To assess whether CNNs are an appropriate tool to be used on interaction maps, we first have to analyze what constitutes the topology in an interaction map and why two points are within or outside each others' local area.

Consider three cells of an interaction map E, created via Equation \ref{eqn:interaction_map}, with coordinates $(x,y)$, $(x,y+1)$, $(x,y+4)$. If we consider point $(x,y)$, with a kernel size of 2, $(x,y+1)$ will be in its local area while $(x,y+4)$ will not. But what is the relation between $y$, $y+1$ and $y+4$? If the embeddings are created with a typical latent factor model (e.g., MF or FM), as done in two of the analyzed papers, the answer is that the latent factors $y$ and $y+1$ are direct neighbors in the embedding vector, while $y$ and $y+4$ are not.

The question now is whether the position of the latent factors has a specific meaning. If we look at typical \MF algorithms, such as \BPRMF or \iALS \cite{hu2008IALS}, we can see that a prediction is computed as follows \cite{rendle2009bprMF}.
\begin{align}
\label{eqn:prediction_bprmf}
\hat{r}_{ui} = p_u^T \cdot q_i = \sum^{K}_{k} p_{u,k} \cdot q_{i,k}
\end{align}
Here, $\hat{r}_{ui}$ is the predicted relevance for user $u$ on item $i$, $p_u$ and $q_i$ are the user and item embedding vectors, and $k$ is the latent factor index.

From Equation \ref{eqn:prediction_bprmf}, we see that the ordering of the latent factors has no impact on the prediction, and the model only requires a biunivocal correspondence between the columns of $P$ and $Q$, regardless of their relative ordering. Such a lack of a \emph{natural ordering} of the latent factors is common to many \MF algorithms including \BPRMF, AsySVD, and \iALS \cite{rendle2009bprMF,hu2008IALS}. 
Only for some techniques, like PureSVD  \cite{Cremonesi:2010:PRA:1864708.1864721}, the latent factors are ordered according to the decreasing singular value they are associated with. Due to this lack of a natural ordering, the specific arrangement of the latent factors is mainly a contingent property and a multitude of equivalent models can be learnt from the same data due to the stochastic nature of the training process.
Each permutation of the ordering of the factors leads to an equivalent MF model but to a different interaction map, which will exhibit different local features.

\begin{figure}[h!t]
\centering
\begin{minipage}[b]{0.45\linewidth}
    \centering
    \subfigure[Vector $u$]{\includegraphics[width=.8\textwidth]{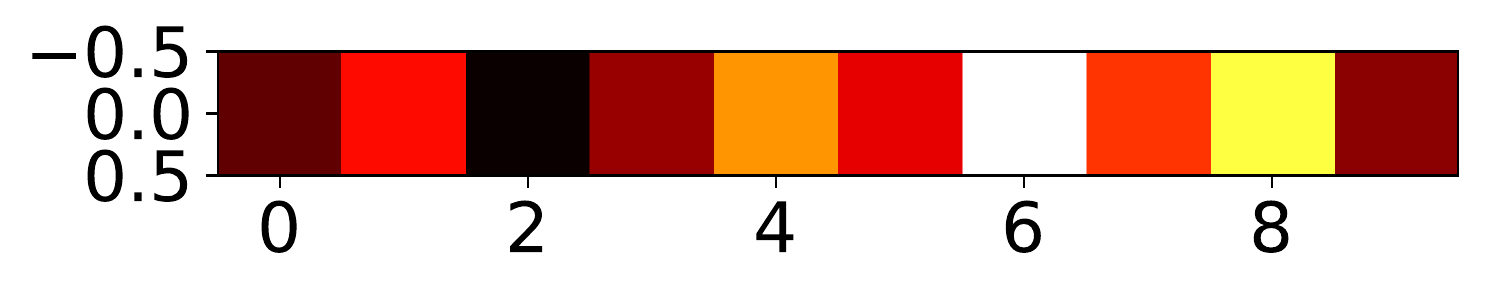}}\quad
    \subfigure[Vector $v$]{\includegraphics[width=.8\textwidth]{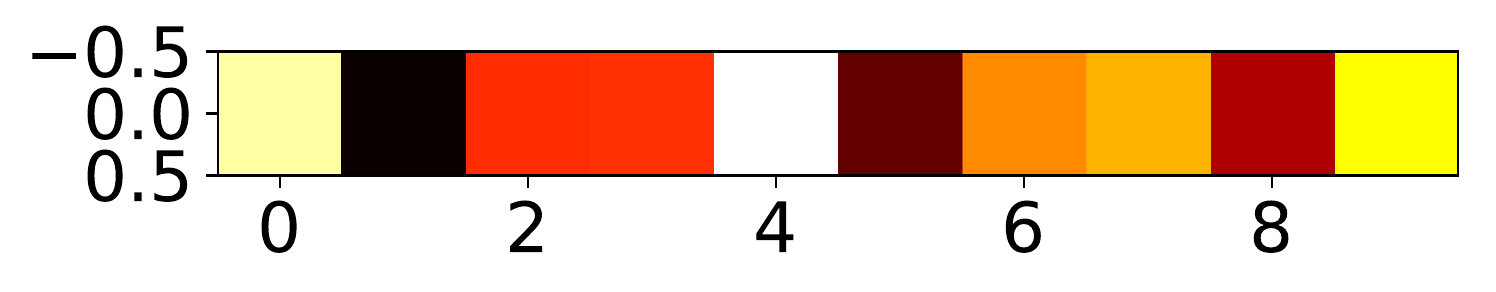}}\quad
    \subfigure[Interaction map]{\includegraphics[width=.8\textwidth]{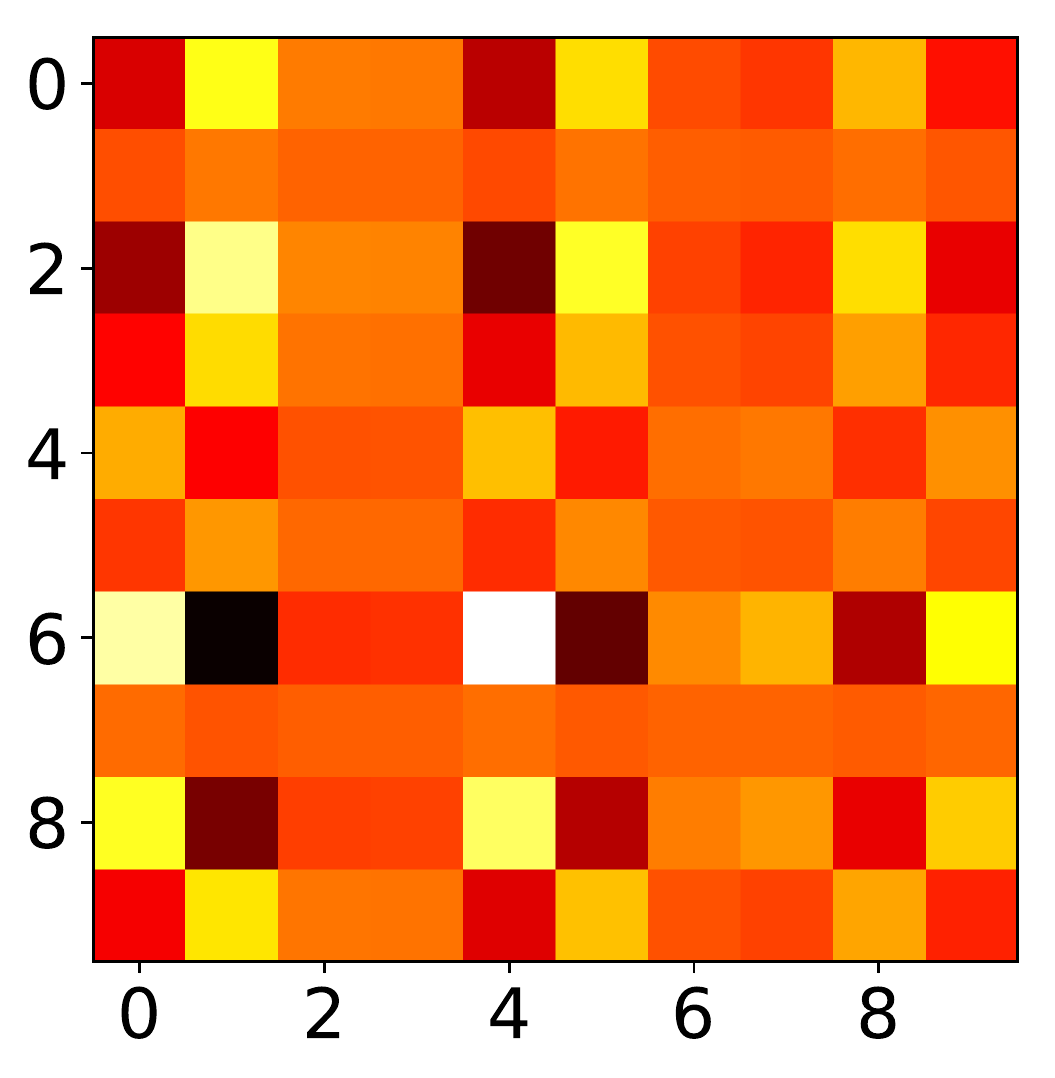}}
\end{minipage}
\begin{minipage}[b]{0.45\linewidth}
    \centering
    \subfigure[Vector $u$]{\includegraphics[width=.8\textwidth]{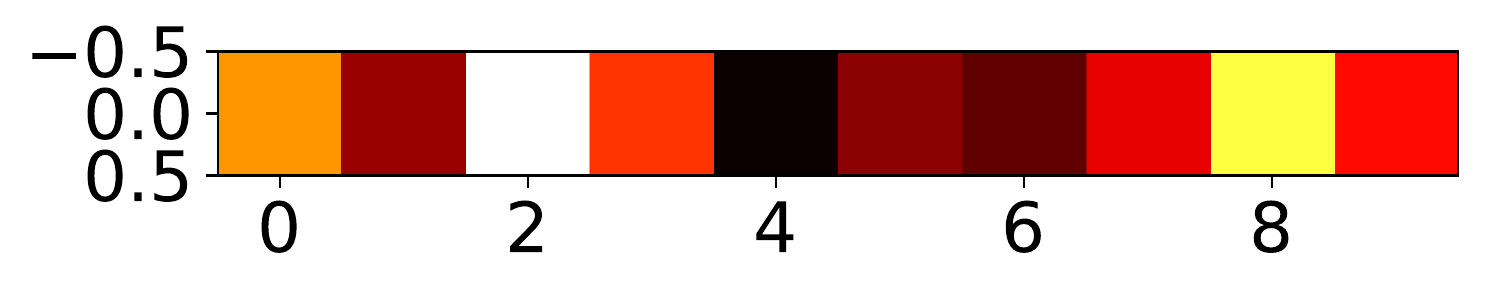}}\quad
    \subfigure[Vector $v$]{\includegraphics[width=.8\textwidth]{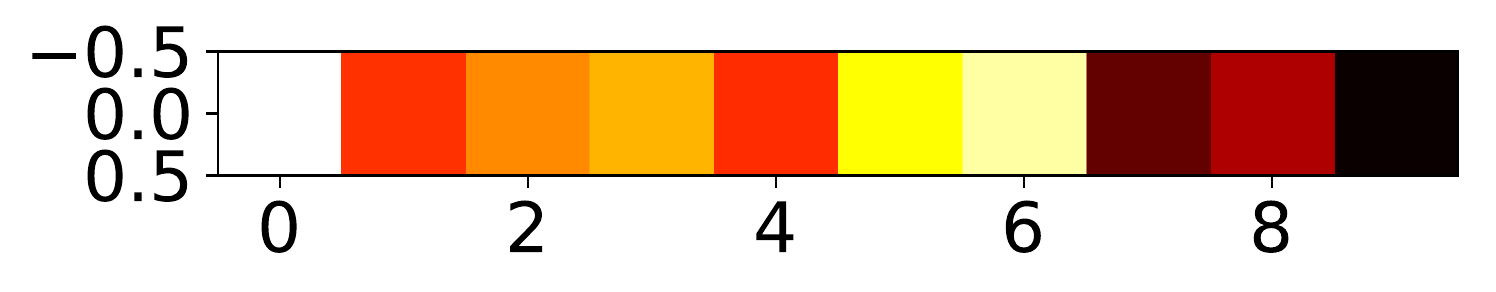}}\quad
    \subfigure[Interaction map]{\includegraphics[width=.8\textwidth]{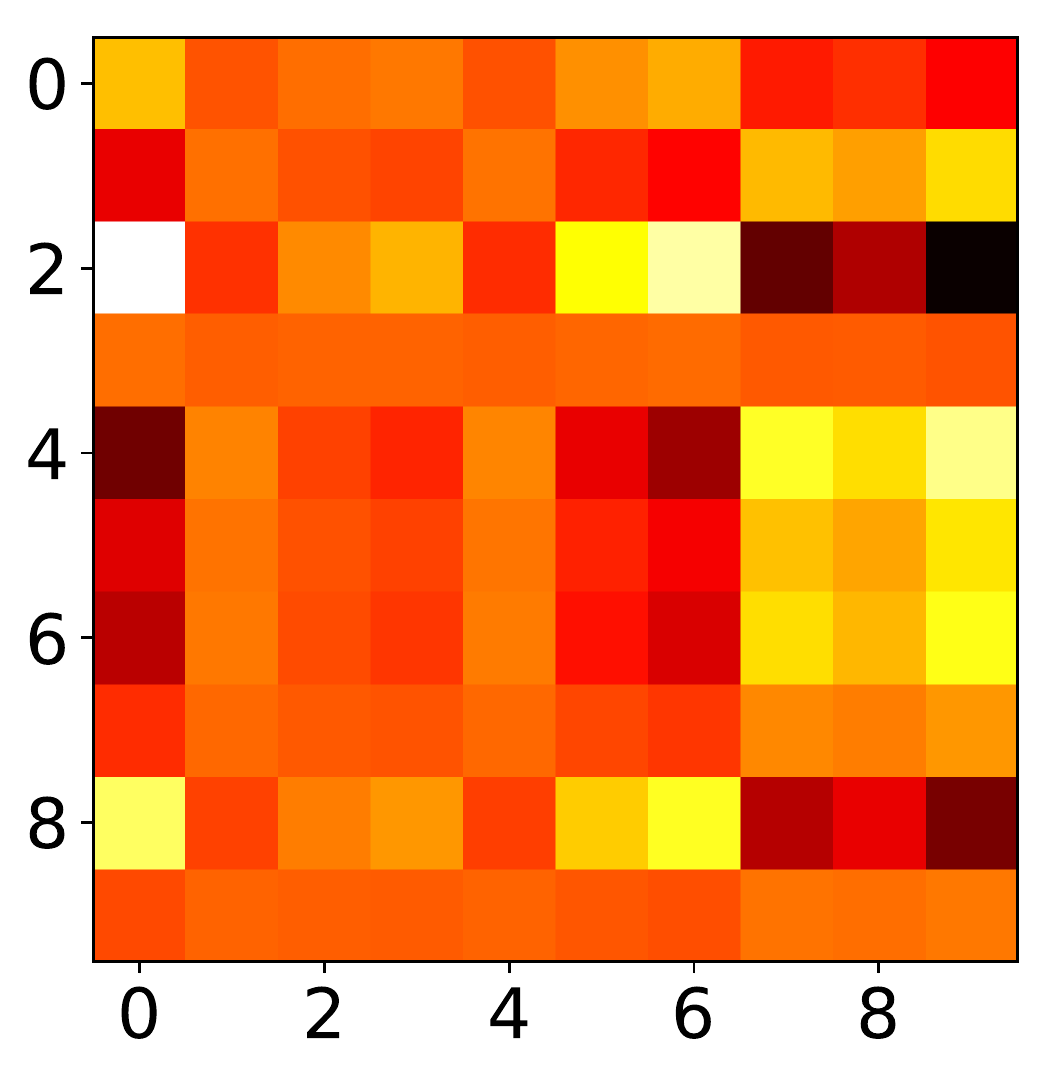}}
\end{minipage}\caption{Effects of permutating the columns of vectors $u$ and $v$ on their resulting outer product (the interaction map).}
\label{fig:heat_map}
\end{figure}

The effects of permutating the columns of the vectors can also be analyzed visually. Consider two randomly created vectors $u$ and $v$ of length 10 (Figures \ref{fig:heat_map}.a and \ref{fig:heat_map}.b) and
two permutations of the same vectors (Figures \ref{fig:heat_map}.d and \ref{fig:heat_map}.e), where darker cells indicate higher values.
It can be easily seen that the interaction map of the original vectors (Figures \ref{fig:heat_map}.c) and the one of the permuted vectors (Figure \ref{fig:heat_map}.f)
do not have any identifiable pattern in common.

This lack of a natural ordering provides evidence that the interaction map is not analogous to an image because it does not exhibit spacial locality (i.e., meaningful local features).

\subsection{Experiment Configurations}
We conducted three types of computational experiments.\footnote{We share the code and data used in our experiments online: \url{https://github.com/MaurizioFD/RecSys2019_DeepLearning_Evaluation}} Two of them, discussed in Section \ref{sec:varying-input-topology} and Section \ref{sec:ablation-studies}, were designed to provide evidence that convolutions over the user-item interaction map do not lead to the claimed effects. The third, discussed in Section \ref{sec:comparison-with-baselines}, shows that the analyzed CNN-based methods are consistently outperformed by existing non-neural techniques, which points to a problem in how the baselines were selected and optimized in the original works.

In all our studies, we used the same experimental designs as in the original papers. In particular, we used the original code, data, data splits, as well as hyperparameters that were provided by the authors.\footnote{This is appropriate because the optimization problem is unchanged and the train-test split remains identical.} To determine the number of epochs, which is not usually reported, we apply early-stopping,
see also \cite{Ferraridacrema2019}. The training is stopped if the recommendation quality does not improve for 5 consecutive evaluation steps. The evaluation setup for each method was as follows.

\begin{description}
  \item[\ConvNCF] was evaluated on a dataset from Yelp.
  The algorithm code and the data split for the Yelp data was published by the authors, but not the preprocessing code. We preprocessed the data based on the information in the paper. \BPRMF was used to pretrain the latent factors.
  \item [\CFM] was tested on a music dataset from Last.fm. 
  The code and the preprocessed data split for Last.fm were provided by the authors. The latent factors were pretrained using a Factorization Machine.
  \item [\CoupledCF]  was tested on the Movielens1M dataset, which also contains side information about users and items. \CoupledCF, as stated above, does not use pretrained models but learnable embeddings.
\end{description}

Note that in the original papers additional evaluations with similar outcomes were reported for alternative datasets. For the purpose of this present study, which aims to show that CNNs in principle cannot work as claimed, it is sufficient to provide one counterexample. Therefore, we limit our discussions to one of the datasets that was used in the original paper.

In all original papers the authors use a leave-one-out evaluation procedure. In two cases a number of randomly sampled negative items (e.g., 99 for \CoupledCF) were ranked with the true positive. The Hit Rate (HR) and the NDCG are used as evaluation measures in all papers, using different cut-off lengths.\footnote{Due to the leave-one-out procedure, other metrics like Recall, Precision and F1 are linearly correlated to HR.} In our evaluation we applied, for each algorithm, the exact same evaluation setting as described in the original paper.

\subsection{Varying the Input Topology}
\label{sec:varying-input-topology}
In our first experiment, we varied the topology of the inputs (i.e., the order of the latent factors) which are fed to the CNN. Given our theoretical considerations from Section \ref{subsec:theoretical-considerations}, altering the topology should have virtually no impact on the model quality because the topology cannot provide any relevant information. To empirically validate this consideration, we designed the following experiment for the approaches that use pretrained embeddings (\ConvNCF and \CFM).

First, we pre-train the embeddings as done in the original articles, i.e., using either \BPRMF or FM.
We then create 20 equivalent pre-trained models by randomly permutating the positions of the latent factors. Each permutation is applied consistently on the user and item latent factors in a way that the latent factor model remains equivalent.
Each permutation will lead to different interaction maps.
For each of these permutations, a convolution model is trained, and the quality of each resulting model is evaluated based on the measures used in the original papers (HR and NDCG) at cutoff 10.

\begin{table}[h!t]
    \centering
    \begin{tabular}{l|ccc}
    \toprule
	            & NDCG 	& HR 	\\
    \midrule
    \BPRMF 	    &0.1576 $\pm$ 0.0000	&0.2966 $\pm$ 0.0000 \\
    ConvNCF 	&0.1623 $\pm$ 0.0008	&0.3052 $\pm$ 0.0019	\\
    \midrule
    \midrule
    FM 	        &0.1230 $\pm$ 0.0000		&0.2234 $\pm$ 0.0000\\
    CFM     	&0.1730 $\pm$ 0.0398    	&0.3155 $\pm$ 0.0724	\\
	\bottomrule
   	\end{tabular}
    \caption{Averaged performance results and standard deviations obtained for 20 permutations of the interaction maps.
    }
    \label{tab:permutation_experiment}
\end{table}

The results of this experiments are reported in Table \ref{tab:permutation_experiment}. 
We both report 
\begin{enumerate*}[label=\textit{(\roman*)}]
  \item the results for the model without the CNN layer (\BPRMF and FM) and
  \item the results for the full model (\ConvNCF and \CFM).
\end{enumerate*}
 For each algorithm, we report the averaged accuracy and the standard deviation resulting from the 20 permutations.
The following observations can be made.

\begin{itemize}
  \item For the ``plain'' \BPRMF and FM models, the standard deviation is zero, as expected by definition from Equation (\ref{eqn:prediction_bprmf}). However, also for the CNN-based models, the standard deviation is almost zero. This confirms that changing the input topology has no relevant impact on the results, indicating that the order of the latent factors, as expected, does not matter and no local features can exist in the interaction map. Note that statistical tests like the t-test cannot be applied when the variance is zero.
  \item The CNN models show improved accuracy over the pretraining models. For the \ConvNCF method, the gains are tiny, i.e., we could not reproduce in our experiments that the CNN adds much value.\footnote{See also \cite{DBLP:journals/corr/abs-1905-01395,Ferraridacrema2019} on related problems of reproducing reported gains in the recommender systems literature.} For the \CFM model, which applies a more complex preprocessing step, improvements over the plain FM model \emph{could} be reproduced. These gains, however, cannot be attributed to the fact that the interaction maps can be considered as images (given the irrelevance of the ordering of the ``pixels'').
\end{itemize}

To put these observed gains into perspective, we, as mentioned above, executed additional experiments in which we compared the performance of \CFM and other methods with existing non-neural techniques, see Section \ref{sec:comparison-with-baselines}, The results are in line with previous observations in \cite{Ferraridacrema2019} and show that the gains happen at a performance level that is largely below the performance of a traditional ItemKNN method \cite{sarwar2001item} (see Table \ref{tab:cfm-lastfm}).

\begin{table*}[h]
    \centering
    \begin{tabular}{ll|cc|cc}
    \toprule
    \multirow{2}{*}{Algorithm}  &  \multirow{2}{*}{Mode} & \multicolumn{2}{c}{Ablation Study 1}   \vline  & \multicolumn{2}{c}{Ablation Study 2} \\
	              &          & NDCG 	& HR   & NDCG 	& HR 	\\
    \midrule
    ConvNCF & full 	        &0.1623 $\pm$ 0.0008	&0.3052 $\pm$ 0.0019	&0.1623 $\pm$ 0.0008	&0.3052 $\pm$ 0.0019\\
    \midrule
    ConvNCF & element-wise 	&0.1622 $\pm$ 0.0008	&0.3051 $\pm$ 0.0016	&0.1632 $\pm$ 0.0012	&0.3068 $\pm$ 0.0015\\
    ConvNCF & correlations    &0.0193 $\pm$ 0.0076	&0.0403 $\pm$ 0.0150	&0.1522 $\pm$ 0.0013	&0.2900 $\pm$ 0.0020\\
    \midrule
    \midrule
    CFM & full 		        &0.1730 $\pm$ 0.0398	    &0.3155 $\pm$ 0.0724    &0.1730 $\pm$ 0.0398	&0.3155 $\pm$ 0.0724	\\
    \midrule
    CFM & element-wise 		&0.1730 $\pm$ 0.0398	    &0.3155 $\pm$ 0.0724	&0.1805 $\pm$ 0.0034	&0.3292 $\pm$ 0.0062    \\
    CFM & correlations 	    &0.0015 $\pm$ 0.0003	    &0.0032 $\pm$ 0.0008	&0.0011 $\pm$ 0.0001	&0.0019 $\pm$ 0.0002    \\
    \midrule
    \midrule
    CoupledCF & full 		    &0.5272 $\pm$ 0.0491		&0.7865 $\pm$ 0.0470	&0.5272 $\pm$ 0.0491	&0.7865 $\pm$ 0.0470\\
    \midrule
    CoupledCF & element-wise 	&0.5404 $\pm$ 0.0631		&0.7744 $\pm$ 0.0994	&\textbf{0.5763 $\pm$ 0.0059}	&\textbf{0.8243 $\pm$ 0.0071}\\
    CoupledCF & correlations 	&0.5137 $\pm$ 0.0903		&0.7822 $\pm$ 0.0659	&0.5503 $\pm$ 0.0343	&0.7978 $\pm$ 0.0391\\
	\bottomrule
  	\end{tabular}
    \caption{Results of the two ablation studies. \emph{Ablation Study 1} evaluates the contribution of each component of the interaction map to a model trained on the full map. \emph{Ablation Study 2} evaluates the contribution of training on different parts of the interaction map. Significant improvements over the full map are printed in bold.
    }
    \label{tab:ablation_experiment}
\end{table*}

\subsection{Ablation Studies}
\label{sec:ablation-studies}
Despite being on a low performance level, the results shown in Table \ref{tab:permutation_experiment} indicate that the CNN layer, at least in one of the cases, seems to have at least some positive effect on the overall performance.
According to our theoretical considerations, these gains cannot stem from leveraging correlations in the embeddings as claimed in the papers, but might be merely the result of adding a neural network layer to the pretraining model, which acts as a universal approximator. We conducted two types of ablation studies to further investigate this question.

\paragraph{Ablation Study 1}
In order to measure how much the CNN model has learned to represent the embeddings correlation, we designed a novel type of ablation study, which was not part of the original papers.
We started from the models previously trained on the full interaction map, but we computed the recommendations using only certain interaction map components. For \CoupledCF, which does not use pretrained embeddings, we trained and evaluated the model on 20 random train-test splits.

Remember that the correlations in the embeddings are represented by the elements of the interaction map (Equation \ref{eqn:interaction_map}) that are not on the main diagonal. 
The following configurations were tested:
\begin{itemize}
  \item \emph{full} : This corresponds to the original setting.
  \item \emph{element-wise}: Only the element-wise products (i.e., main diagonal elements) are considered. 
          \begin{equation*}
        e_{x,y} = \left\{
         \begin{array}{@{}l@{\thinspace}l}
           p_{u,x} \cdot q_{i,y}  &\quad \text{if } x = y\\
           0                        &\quad \text{otherwise.}\\
         \end{array}
        \right.
        \end{equation*}
  \item \emph{correlations}: Only the embeddings correlations (i.e., off-diagonal elements) are used.
      \begin{equation*}
    e_{x,y} = \left\{
     \begin{array}{@{}l@{\thinspace}l}
       p_{u,x} \cdot q_{i,y}  &\quad \text{if } x \neq y\\
       0                        &\quad \text{otherwise.}\\
     \end{array}
    \right.
    \end{equation*}
\end{itemize}

The results of this ablation study are reported in Table \ref{tab:ablation_experiment}. They show that there is no statistical difference\footnote{The statistical significance of the difference between the observed results with $\alpha$=0.05 was verified using a paired t-test if the values were normally distributed; and a Wilcoxon signed-rank test otherwise. To assess if the values of the metrics are normally distributed we used both Shapiro-Wilk and Kolmogorov-Smirnov tests.} between the models evaluated using the full interaction map and when only using the element-wise product (diagonal elements).

In other words, the off-diagonal correlation elements are not contributing anything to the overall performance.
Interestingly, when the recommendations are computed using only the embeddings correlation, the observed results for \ConvNCF and \CFM are lower by at least an order of magnitude. This suggests both models have in some ways \emph{learned to ignore the embedding correlations}.
For \CoupledCF, the accuracy obtained with the embeddings correlation is similar to the full interaction map. Therefore, although the model has learned to use the embeddings correlation, this did not prove to be beneficial for the model's quality.
Overall, the results clearly indicate that the convolutional models are not learning to represent the embeddings correlation when these are pre-trained (\idest \ConvNCF and \CFM). They also do not benefit from them even when they are learnable (\CoupledCF), which is in direct contradiction to what was stated in the original articles.

Remember that in the original articles a similar ablation study was not present. The contribution of the embeddings correlation to the convolution model was therefore never directly measured.

\paragraph{Ablation Study 2}
In \emph{Ablation Study 1}, we observed that training the model on the full interaction map resulted in the embeddings correlation not contributing to improve the performance over the simple element-wise product.

While in \emph{Ablation Study 1} we trained the model on the full interaction map, in \emph{Ablation Study 2}, we isolate the different components of the interaction map at an earlier stage, i.e., during \emph{the training phase}. Therefore, we do not train the network on the full map as in \emph{Ablation Study 1}, but instead we train the model on specific components of the map.

In this new experiment, different models are therefore trained from scratch, using only the interaction map component associated with a given configuration (i.e., \emph{full-map}, \emph{element-wise}, \emph{correlations}).\footnote{As discussed by Rendle et al. \cite{ncfvsmf2020} in the element-wise configuration it should be trivial for a simple CNN to learn the dot product of the embeddings, since the dot product is equal to the diagonal of the outer product.} As a result, a model trained only on the element-wise product will never observe embeddings correlations and vice versa. This allows us to measure how much of each component the convolution algorithm learns to model when the other is not present.

The results of \emph{Ablation Study 2} are also reported in Table \ref{tab:ablation_experiment}. Since the training data (i.e., interaction map) fed to the CNN in the two ablation studies are different, it is expected that the absolute values of the measurements are different. However, we can again observe that the results obtained when training the convolution model on the full interaction map and on the element-wise product are not different to a statistically significant extent for \ConvNCF and \CFM. The convolution operation is therefore not leveraging the embedding correlations in any effective way. As a result, these correlations can be discarded entirely during the training phase without degrading the model's performance. Remarkably, for \CoupledCF we can observe that training the model on the element-wise product alone results in significantly better results than when using the full map.
Remember that \CoupledCF does not use pretrained embeddings but learns them during the training process, while \ConvNCF and \CFM, use pre-trained embeddings instead. Our result indicate that the additional parameter space provided by the learnable embedding correlations even seems to introduce noise, effectively harming the model quality.

Interestingly and differently from \emph{Ablation Study 1}, we can see for \ConvNCF and \CoupledCF that training the convolution on the embeddings correlation elements alone yields results that are very close to those obtained when using the full interaction map.
This suggests that the pretrained embeddings in \ConvNCF do indeed carry some information that can, to an extent, be modeled. Similarly, \CoupledCF can learn some correlations, although with a rather high standard deviation. However, the CNN models are, as demonstrated through \emph{Ablation Study 1}, not able to leverage correlations to improve the accuracy obtained on the element-wise product alone.

There may be different reasons for this. First, it might be that the convolution on the full map was only able to model information that was redundant and already captured by the element-wise model. An alternative explanation is that the CNN model did not succeed in hybridizing these two pieces of information in a synergistic way, i.e., it only learned to select the best-performing or the less noisy one.

\subsection{Comparison with Non-Neural Baselines}
\label{sec:comparison-with-baselines}
In all original papers investigated here, the claim is made that the proposed CNN-based algorithm is able to outperform the state-of-the-art. This claim is also used as demonstration that all models are effectively able to leverage embeddings correlations, which, as we showed before, is not the case. Recent research has found several instances of works where similar claims were possible only due to methodological issues such as the choice of weak baselines, the lack of proper optimization of the baselines, or information leakage from the test data
\cite{lin2019recantation,DBLP:journals/corr/abs-1905-01395,Ferraridacrema2019}. In particular, two of the algorithms we analyze here (\ConvNCF and \CoupledCF) have been reported in \cite{Ferraridacrema2019troubling} to be not competitive against simple baselines. We could replicate the performance results reported in \cite{Ferraridacrema2019troubling} for \ConvNCF and \CoupledCF. Furthermore, we have conducted additional experiments of the same form for \CFM, for which such an analysis was missing so far.

Like in Ferrari Dacrema et al. \cite{Ferraridacrema2019,Ferraridacrema2019troubling}, we compared all CNN-based algorithms to the same set of known non-neural, adequately optimized baseline algorithms. We used the evaluation framework shared by \cite{Ferraridacrema2019,Ferraridacrema2019troubling}, and extended the framework with an implementation of the \CFM method which was made publicly available by the authors of the method.\footnote{The results obtained for \ConvNCF and \CoupledCF refer to the same train-test split and hyperparameter setting and are, therefore, perfectly reproducing the results reported in \cite{Ferraridacrema2019troubling}.}

\subsubsection{Baseline algorithms}
The baseline algorithms we report here are a subset of those used in \cite{Ferraridacrema2019troubling}, and they represent algorithms of different families.
\begin{description}
  \item[TopPopular] A non-personalized model recommending the most popular items.
  \item[ItemKNN] An item-based nearest neighbor model \cite{sarwar2001item} using cosine similarity and shrinkage.
  \item[UserKNN] A user-based nearest neighbor model \cite{Resnick:1994:GOA:192844.192905} using cosine similarity and shrinkage.
  \item[\palpha] A graph-based model implementing a random-walk between user and item nodes \cite{cooper2014P3alpha}. The method is equivalent to a KNN item-based CF algorithm, with the similarity matrix being computed as the dot-product of the probability vectors.
  \item[\pbeta] A version of \palpha which involves a reranking step \cite{paudel2017Rp3beta}.
  \item[PureSVD] A matrix factorization approach based on the traditional SVD decomposition \cite{Cremonesi:2010:PRA:1864708.1864721}.
  \item[Sparse Linear Models (SLIM)] An item-based recommendation model that learns the similarity between items via linear regression \cite{ning2011SLIM}. In our work, we use the more scalable variant proposed in \cite{levy2013SLIM_ElasticNet}.
  \item[Implicit Alternating Least Squares (\iALS)] A matrix factorization model for implicit feedback datasets proposed in \cite{hu2008IALS}. In the \iALS method, the implicit feedback signals are transformed into confidence values.
\end{description}

\subsubsection{Hyperparameter optimization}
In order to optimize the hyperparameters of the baseline methods we create a validation split from the train data, by applying the same splitting procedure that was used to create the test data. We use a Bayesian search \cite{antenucci2018artist,NIPS2014_5324,Freno:2015:ORM:2783258.2788579}, available as part of the Scikit-Optimize\footnote{\url{https://scikit-optimize.github.io/}} package, exploring 50 hyperparameter configurations, with the first 15 used as an initial random initialization. Once hyperparameter values were found that optimize the recommendation accuracy on the validation data, we train the model again on the union of train and validation data and report the recommendation accuracy on the test data.
Each baseline algorithm has a different set of hyperparameters. The complete list of these parameters as well as their range and distribution is the same as in reported in \cite{Ferraridacrema2019troubling}.

\subsubsection{Results}
The result of this comparison against simple baselines can be observed in Table \ref{tab:convncf-yelp} (\ConvNCF), Table \ref{tab:coupledcf-movielens} (\CoupledCF) and Table \ref{tab:cfm-lastfm} (\CFM).
For all of these algorithms it was possible to reproduce both the experimental setting (i.e., the source code developed by the original authors and at least one dataset were available) and the numerical results reported in the original paper.

However, as it is possible to observe, we could not confirm that any of the algorithms is able to outperform the state-of-the-art.
\begin{itemize}
  \item \ConvNCF on the Yelp dataset is outperformed by all baselines, sometimes by more than 10\%. Similar observations were made for the Gowalla dataset, which served as a second dataset in the original paper.
  \item  \CoupledCF on Movielens1M is able to achieve recommendation accuracy results that are competitive with neighborhood-based and graph-based baselines. It is however not competitive with known techniques based on matrix factorization or linear regression. The results are similar for the Tafeng dataset used in the original paper.
  \item Lastly, \CFM achieves surprisingly poor recommendation accuracy, sometimes only reaching half the level of the baselines for the Last.fm dataset. For the second dataset that was used to evaluate \CFM in \cite{ijcai2019CFM}, we could not reproduce the original results because the code shared by the authors did not execute correctly\footnote{We contacted all the authors to resolve the issue, but without success.}.
\end{itemize}

Overall, our results indicate that the analyzed CNN-based algorithms, despite their computational complexity\footnote{Even on high-end GPUs, the computation time need to fit the CNN models including early stopping on a typical dataset is about 8 hours for \CoupledCF, 17 hours for \CFM and 24 hours for \ConvNCF.}, are actually not able to outperform comparably simple and long-known baselines and to advance the state-of-the-art. While the use of an additional CNN layer may have some positive effects in some cases, these effects are
\begin{enumerate*}[label=\textit{(\roman*)}]
  \item not the result of CNNs capturing correlations in the user-item interaction maps,
  \item not sufficient to raise the performance level of the CNN-based methods over the state-of-the-art.
\end{enumerate*}

\begin{table}[h!t]
    \caption{Experimental results for ConvNCF for the Yelp dataset.}
    \label{tab:convncf-yelp}
    \small
    \centering
    \begin{tabular}{l|cc|cc|cc}
    \toprule
    & \multicolumn{2}{c}{@5}  \vline & \multicolumn{2}{c}{@10}  \vline & \multicolumn{2}{c}{@20} \\
    & HR 	& NDCG 	& HR 	& NDCG 	& HR 	& NDCG 	\\
    \midrule
    TopPopular	&0.0817	&0.0538	&0.1200	&0.0661	&0.1751	&0.0799	\\
    \midrule
    UserKNN CF     &  \textbf{0.2068} &  \textbf{0.1355} &  \textbf{0.3126} &  \textbf{0.1695} &           0.4401 &  \textbf{0.2017} \\
    ItemKNN CF	&\textbf{0.2521}	&\textbf{0.1686}	&\textbf{0.3669}	&\textbf{0.2056}	&\textbf{0.4974}	&\textbf{0.2385}	\\
    \midrule
    \palpha &\textbf{0.2146}	&\textbf{0.1395}	&\textbf{0.3211}	&\textbf{0.1737}	&0.4442	&\textbf{0.2049}	\\
    \pbeta	&\textbf{0.2202}	&\textbf{0.1431}	&\textbf{0.3323}	&\textbf{0.1793}	&\textbf{0.4667}	&\textbf{0.2132}	\\
    \midrule
    SLIM	&\textbf{0.2330}	&\textbf{0.1535}	&\textbf{0.3475}	&\textbf{0.1904}	&\textbf{0.4799}	&\textbf{0.2238}	\\
    PureSVD	&\textbf{0.2011}	&\textbf{0.1307}	&0.3002	&\textbf{0.1626}	&0.4238	&0.1938	\\
    \iALS	&\textbf{0.2048}	&\textbf{0.1348}	&\textbf{0.3080}	&\textbf{0.1680}	&0.4319	&\textbf{0.1993}	\\
    \midrule
    ConvNCF	&0.1947	&0.1250	&0.3059	&0.1608	&0.4446	&0.1957	\\
    \bottomrule
  	\end{tabular}

\end{table}

\begin{table}[h!t]
    \caption{Experimental results for CoupledCF for the MovieLens1M dataset.}
    \label{tab:coupledcf-movielens}
    \small
    \centering
    \begin{tabular}{l|cc|cc|cc}
    \toprule
    {} & \multicolumn{2}{c}{@ 1} & \multicolumn{2}{c}{@ 5} & \multicolumn{2}{c}{@ 10} \\
    {} &               HR &             NDCG &               HR &             NDCG &               HR &             NDCG \\
    \midrule
    TopPopular                    &           0.1593 &           0.1593 &           0.4217 &           0.2936 &           0.5813 &           0.3451 \\
    \midrule
    UserKNN CF         &           0.3540 &           0.3540 &           0.6884 &           0.5324 &           0.8060 &           0.5704 \\
    ItemKNN CF         &           0.3305 &           0.3305 &           0.6682 &           0.5080 &           0.7940 &           0.5488 \\
    \midrule
    \palpha                   &           0.3316 &           0.3316 &           0.6543 &           0.5031 &           0.7687 &           0.5402 \\
    \pbeta                   &           0.3464 &           0.3464 &           0.6743 &           0.5198 &           0.7959 &           0.5591 \\
    \midrule
    SLIM           &  \textbf{0.3906} &  \textbf{0.3906} &  \textbf{0.7116} &  \textbf{0.5625} &  \textbf{0.8315} &  \textbf{0.6014} \\
    PureSVD                   &  \textbf{0.3735} &  \textbf{0.3735} &  \textbf{0.7088} &  \textbf{0.5522} &           0.8132 &  \textbf{0.5861} \\
    \iALS                      &  \textbf{0.3816} &  \textbf{0.3816} &  \textbf{0.7121} &  \textbf{0.5581} &           0.8200 &  \textbf{0.5933} \\
    \midrule
    CoupledCF                 &           0.3522 &           0.3522 &           0.7018 &           0.5374 &           0.8247 &           0.5775 \\
    \bottomrule
    \end{tabular}

\end{table}

\begin{table}[h]
    \caption{Experimental results for CFM for the Last.fm dataset.}
    \label{tab:cfm-lastfm}
    \small
    \centering
    \begin{tabular}{l|cc|cc|cc}
    \toprule
     & \multicolumn{2}{c}{@5}  \vline & \multicolumn{2}{c}{@10}  \vline & \multicolumn{2}{c}{@20} \\
		& HR 	& NDCG 	& HR 	& NDCG 	& HR 	& NDCG 	\\
		\midrule
    TopPopular	&0.0016	&0.0009	&0.0023	&0.0011	&0.0033	&0.0014	\\
    \midrule
    UserKNN CF	&\textbf{0.5964}	&\textbf{0.4527}	&\textbf{0.6715}	&\textbf{0.4773}	&\textbf{0.7032}	&\textbf{0.4855}	\\
    ItemKNN CF	&\textbf{0.5975}	&\textbf{0.4425}	&\textbf{0.6776}	&\textbf{0.4689}	&\textbf{0.7070}	&\textbf{0.4764}	\\
    \midrule
    \palpha	&\textbf{0.6327}	&\textbf{0.4929}	&\textbf{0.6744}	&\textbf{0.5066}	&\textbf{0.7014}	&\textbf{0.5135}	\\
    \pbeta	&\textbf{0.5896}	&\textbf{0.4458}	&\textbf{0.6756}	&\textbf{0.4739}	&\textbf{0.7071}	&\textbf{0.4821}	\\
    \midrule
    SLIM	&\textbf{0.6674}	&\textbf{0.5169}	&\textbf{0.6972}	&\textbf{0.5267}	&\textbf{0.7102}	&\textbf{0.5300}	\\
    PureSVD	&\textbf{0.4026}	&\textbf{0.3117}	&\textbf{0.4891}	&\textbf{0.3397}	&\textbf{0.5652}	&\textbf{0.3590}	\\
    \iALS	&\textbf{0.6110}	&\textbf{0.4811}	&\textbf{0.6735}	&\textbf{0.5017}	&\textbf{0.7033}	&\textbf{0.5093}	\\
    \midrule
    CFM	&0.2241	&0.1485	&0.3338	&0.1839	&0.4661	&0.2173	\\
    \bottomrule
  	\end{tabular}
\end{table}

\section{Conclusions}
\label{sec:conclusions}
In this work, we analyzed recently published articles using CNNs on the interaction maps obtained from user and item embeddings. We argued that the original articles lacked a proper discussion on two crucial claims they made: 
\begin{enumerate*}[label=\textit{(\roman*)}]
\item the analogy of embeddings and images, and 
\item the ability of CNNs to model embeddings correlations.
\end{enumerate*}

Our work has shown both through theoretical considerations and through empirically studies that embeddings do not share the topological properties of images. The use of CNNs is therefore not well justified since the embeddings interaction map does not exhibit feature locality and translation invariance, hence it is not analogous to an image. Moreover, we have shown that CNNs, as opposed to what was claimed in the original articles, are insensitive to the embeddings correlation and fail to improve over a model only using the element-wise product. Furthermore, we have compared the CNN based algorithms against a set of established and well optimized non-neural baselines. This was done using the same experimental setup as reported in the original papers. We could show that the proposed complex CNN algorithms were not able to outperform the state-of-the art.

Overall, while we do not argue that convolution cannot be applied on embeddings, we stress that a deeper understating and theoretical analyses of the semantics that new approaches are claimed to leverage are essential to obtain reliable progress in this field. Similarly, claims regarding the improved modelling capacity of an algorithm cannot be based simply upon its ability to outperform a set of baseline algorithms and datasets whose choice is not well justified.
Instead, these aspects should be directly verified via specifically designed experiments.

Ultimately, our work also puts forward a new research question, which previously did not receive much attention. Specifically, the question is how to create an interaction map that captures potentially existing correlations in the data in its topology, such that CNNs can be successfully leveraged on these embeddings interaction maps.

\balance